\documentclass[letterpaper, 10pt, conference]{IEEEconf}  

\IEEEoverridecommandlockouts                              
\usepackage{amsmath}
\usepackage{amssymb}
\usepackage{amsthm}
\usepackage{algorithmic}
\usepackage{algorithm}
\usepackage{mathtools}
\usepackage{xcolor}
\usepackage{makecell}
\usepackage{graphicx}
\usepackage{marginnote}
\usepackage{cite}
\newtheorem{remark}{Remark}
\newtheorem{assumption}{Assumption}
\newtheorem{thm}{Theorem}

\newtheorem{definition}{Definition}

\newcommand{\nc}{\mathrm}
\newcommand{\bm}{\boldsymbol}
\newcommand{\n}{\mathbf}
\newcommand{\cm}{\mathcal}
\newcommand{\vu}{\mathbf{u}}
\newcommand{\xp}{\hat{x}^p}
\newcommand{\xq}{\hat{x}^q}
\newcommand{\xpxq}{p\rightarrow q}
\usepackage[bottom]{footmisc}

\renewcommand{\qed}{\hfill\blacksquare}

\usepackage{version}

\usepackage{hyperref}

\title{\LARGE \bf
Collaborative learning model predictive control for repetitive tasks
}

\author{Chanfreut, P.,  Maestre, J. M.,  Camacho, E. F.,~and~Borrelli, F.
\thanks{P. Chanfreut, J. M. Maestre and E. F. Camacho are with the Department
of Systems and Automation Engineering, University of
Seville,  Spain (e-mails: \texttt{\{pchanfreut,pepemaestre,efcamacho\}@us.es})}
\thanks{F. Borrelli is with the Department of Mechanical Engineering,
University of California at Berkeley, Berkeley, CA 94701 USA (e-mail:
\texttt{fborrelli@berkeley.edu})}
\thanks{This work is supported by the Spanish Training Program for Academic Staff under Grant FPU17/02653, by the Manuel Gayán Buiza Award, by the European Research Council Advanced Grant OCONTSOLAR under Grant SI-1838/24/2018, and by the Spanish MCIN/AEI/10.13039/501100011033 Project C3PO-R2D2 under Grant PID2020-119476RB-I00. Also,  we would like to thank Dr. Filiberto Fele for his feedback regarding the article. 
}
}

\begin{document}

\setlength{\marginparwidth}{1.3cm}

\maketitle
\thispagestyle{empty}
\pagestyle{empty}

\begin{abstract}
This paper presents a cloud-based learning model predictive controller that integrates three interacting components: a set of agents, which must learn to perform a finite set of tasks with the minimum possible local cost; a coordinator, which assigns the tasks to the agents; and the cloud, which stores data to facilitate the agents' learning.  The tasks consist in traveling repeatedly between a set of target states while satisfying input and state constraints. In turn, the state constraints may change in time for each of the possible tasks. To deal with it, different \textit{modes of operation}, which establish different restrictions, are defined. The agents' inputs are found by solving local model predictive control (MPC) problems where the terminal set and cost are defined from previous trajectories. The data collected by each agent is uploaded to the cloud and made accessible to all their peers. Likewise, similarity between tasks is exploited to accelerate the learning process. 
The applicability of the proposed approach is illustrated by simulation results. 
\end{abstract}

\section{Introduction}

In the last decades, model predictive control (MPC) has gained increasing acceptance in both industrial and academic fields, and is now established as a major methodology for dealing with multivariate and constrained systems~\cite{EFC2,qin2003survey}. MPC policies are based on reiterative computations of the sequence of inputs that optimizes the system performance during a future time horizon, thus providing a unique anticipation capacity. 

The increased capability of sensing, computing, and storing data, together with the powerful advances in machine learning techniques, have boosted the application of data-driven methods within the field of MPC~\cite{hewing2020learning}. 
 In this paper, we use the learning model predictive control~(LMPC) formulation presented in~\cite{rosolia2017learning}.  This strategy focuses on systems with a strong repetitive behaviour, such as autonomous racing cars~\cite{rosolia2017autonomous}, and counteracts the inherent finite-horizon nature of MPC controllers. The learning process is based on the systematic design of the terminal set and terminal cost function of the MPC problem by using data. Note that the idea of learning the objective function is also present in other control approaches such as inverse optimal control~\cite{ab2020inverse}. 
 Under the assumption that the model is perfectly known, LMPC controllers \cite{rosolia2017learning} are proved to progressively improve the cost of executing a given task, and closely approximate the solution providing the optimal performance.  Recent efforts have been made to extend the results in~\cite{rosolia2017learning} to decentralized systems and to a more flexible paradigm where the repeated tasks are more varied.  For example, \cite{vallon2020task} presents a task decomposition method where they are defined as aggregations of subtasks in different orders, and~\cite{scianca2020learning} considers the case of
periodically time-varying systems. See also~\cite{zhu2020trajectory}, where a decentralized LMPC for nonlinear multi-agent systems with coupled state constraints is introduced. 

This article presents a cloud-based LMPC for multi-agent systems that repeatedly execute a finite set of tasks, which consist in driving admissibly the agents' state to a certain target point.  The tasks conditions can change with time and thus modify the state restrictions to accomplish them, for instance, consider a mobile robot driving iteratively a road segment with and without the presence of obstacles. To deal with static obstacles, we define different \textit{modes of operation}, which accordingly impose different constraints in the MPC problems.  In addition, a reactive strategy is included to avoid collisions with moving obstacles, which can model for example humans walking. The main novelty of the paper is the use of a data cloud in multi-agent LMPC-based systems as a means to enable collaboration while keeping the computation of the inputs decentralized. In this regard, all agents upload their collected data to a common cloud, so that it can also be used by their peers. Moreover, the data collected when executing a certain task are analyzed to check if they can also be exploited for learning similar ones. The proposed approach guarantees recursive feasibility, asymptotic stability of the target states, and a non-increasing evolution of the cost of executing the same task without moving obstacles. In this respect, under mild assumptions, moving obstacles are proven not to jeopardize persistent feasibility or convergence to the~targets with the proposed controller.

The rest of the paper is organized as follows. Section~II describes the system dynamics and the control goal. Section~III introduces the proposed collaborative learning approach for iterative tasks. Section IV presents its theoretical guarantees. Section V provides the simulation results, and, finally, Section VI presents our conclusions and future work~prospects. 

\section{Problem formulation}\label{sec:problem_formulation}

Consider a set $\cm{N}=\{1,2,\hdots,N_\nc{ag}\}$ of identical agents with linear time-invariant dynamics. Without loss of generality, \textcolor{black}{they will represent mobile agents} navigating in a two-dimensional space. In particular, the continuous-time dynamics of all $i\in \cm{N}$ are given by:\footnote{The approach proposed in this article may be similarly applied to other linear time-invariant systems. However, for the sake of convenience and clarity, we will consider dynamics \eqref{eq:robots_model} throughout the entire manuscript.} 
\begin{equation}\label{eq:robots_model}
  \underbrace{\begin{bmatrix}
    \dot{s}_{\nc{x},i} \\ \dot{s}_{\nc{y},i} \\ \dot{v}_{\nc{x},i} \\ \dot{v}_{\nc{y},i}  
    \end{bmatrix}}_{\dot{x}_i} =   \begin{bmatrix}
    0 & 0 & 1 & 0 \\ 0 & 0 & 0 & 1 \\ 0 & 0 & 0 & 0 \\ 0 & 0 & 0 & 0
    \end{bmatrix}   \underbrace{\begin{bmatrix}
    s_{\nc{x},i} \\ s_{\nc{y},i} \\ v_{\nc{x},i} \\ v_{\nc{y},i}  
    \end{bmatrix}}_{x_i} +  \begin{bmatrix}
    0 & 0 \\ 0 & 0  \\ 1 & 0  \\ 0 & 1
    \end{bmatrix}  \underbrace{\begin{bmatrix}
    a_{\nc{x},i} \\ a_{\nc{y},i} 
    \end{bmatrix}}_{u_i},
\end{equation}
\vspace{-8pt}
\\
\noindent where $s_{\nc{x},i}$, $v_{\nc{x},i}$, and $a_{\nc{x},i}$ are respectively the position, velocity and acceleration of agent~$i$ in the horizontal direction, and $s_{\nc{y},i}$, $v_{\nc{y},i}$, and $a_{\nc{y},i}$  in the vertical direction. 

Hereafter, consider a discrete-time version of \eqref{eq:robots_model} and let us use time index $k$, hence, $x_i(k)$ and $u_i(k)$ denote respectively the state and input of agent $i$ at time instant $k$. Also, define $s_i(k) = [s_{\nc{x},i}(k), s_{\nc{y},i}(k)]^\nc{T}$ and $v_i(k) = [v_{\nc{x},i}(k), v_{\nc{y},i}(k)]^\nc{T}$, and consider set  $ \hat{\cm{X}}=\{\hat{x}^1, \hdots, \hat{x}^p, \hdots, \hat{x}^q, \hdots, \hat{x}^{N_\nc{t}}\}$, 
which contains a finite number $N_\nc{t}$ of \textit{target} states. Finally,  assume  that the initial state of every agent $i \in \cm{N}$ is such that $x_i(0) \in \hat{\cm{X}}$, and $x_i(0)\neq x_j(0)$,  for all $j \in \cm{N}\setminus{\{i\}}$. 
\noindent 

\subsection{Tasks and constraints }\label{sec:agents_tasks}

The goal of the agents is to travel iteratively between the points in $\hat{\cm{X}}$ while optimizing a given performance criterion and satisfying state and input constraints.
\begin{definition}
Task~$\xpxq$ denotes the intent of traveling from $\xp$ to target state $\xq$ in an admissible way, with $\xp, \xq \in \hat{\cm{X}}$.
\end{definition}

\begin{figure}[b]
    \centering
    \includegraphics[scale=0.37,trim={0cm 0 1.5cm 0.2cm},clip]{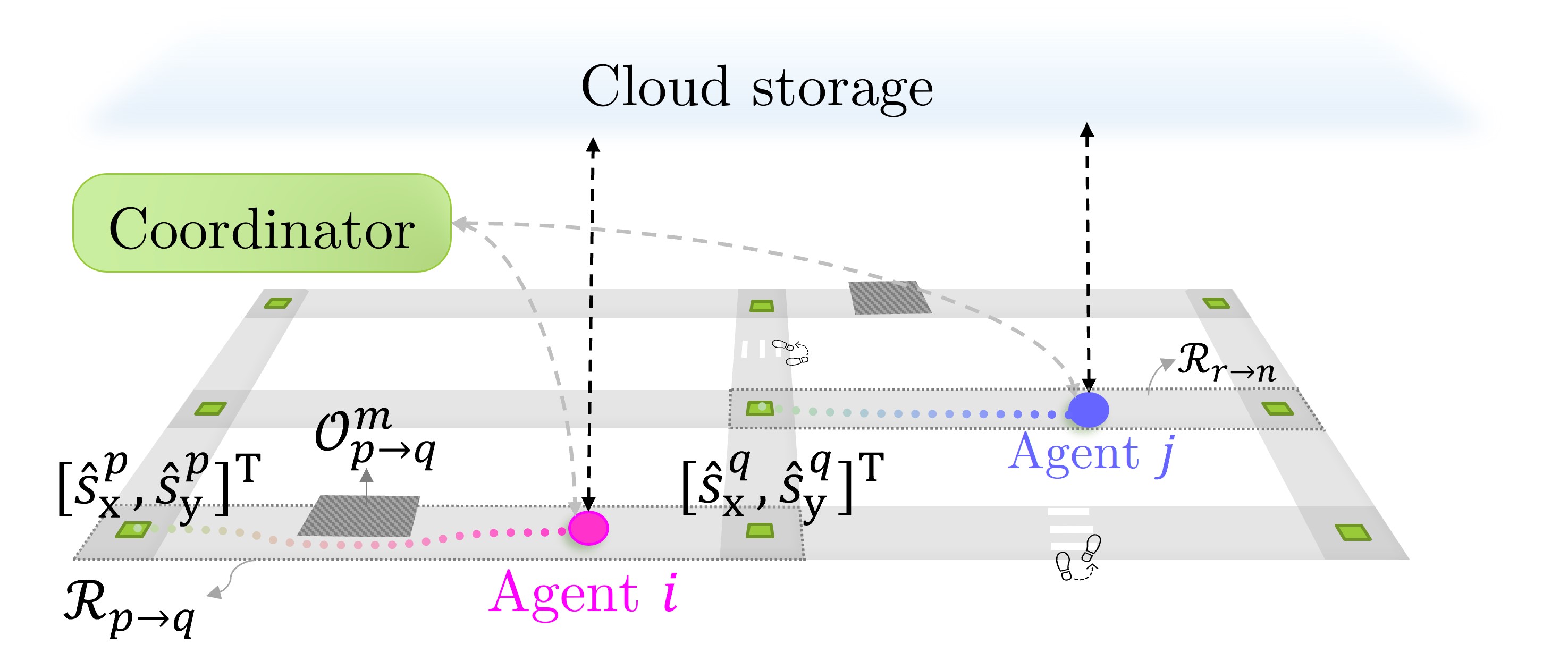}
    \vspace{-0.35cm}
    \caption{Sketch of the system. The tasks consist in traveling admissibly between the target points (green marks). The forbidden areas (dark grey zones) depend on the mode $m$ of each  task and modify the state constraints.} 
    \label{fig:architecture}
\end{figure}

\noindent In this regard, all agents $i \in \cm{N}$ must satisfy 
\begin{subequations}\label{eq:constraints_subsystems}
\begin{align}
     v_i(k) \in \cm{V}&\!=\!\left\lbrace [v_{\nc{x},i},  v_{\nc{y},i}]^\nc{T} \ |  \left\Vert \begin{bmatrix}v_{\nc{x},i}, & \!\! v_{\nc{y},i}\end{bmatrix}\right\Vert_2 \leq v_\nc{max} \right\rbrace\!,  \\  
   u_i(k) \in \cm{U}&\!=\!\left\lbrace [a_{\nc{x},i},  a_{\nc{y},i}]^\nc{T} \ |  \left\Vert \begin{bmatrix}a_{\nc{x},i}, & \!\!\! a_{\nc{y},i}\end{bmatrix}\right\Vert_2 \leq a_\nc{max} \right\rbrace\!,    
\end{align}
\end{subequations}
\noindent for all $k\geq 0$, where $v_\nc{max}$ and $a_\nc{max}$ are the maximum velocity and acceleration. On the other hand,  the constraints on their position depend on the task that is performed. In particular, to perform $\xpxq$, the agents should travel along a path represented by set  $\cm{R}_{\xpxq}$ (see Fig.~\ref{fig:architecture}). Moreover, certain zones of~$\cm{R}_{\xpxq}$ may become non-transitable for a number of time steps, e.g., due to the presence of static obstacles. The latter shrinks the set of
admissible positions, thus modifying the state constraints. It is assumed that these changes can be classified into a finite number of \textit{modes of operation} for each task. In this regard, consider the following:

\begin{definition}
The set of admissible states of any agent~$i$ when performing task $\xpxq$ in mode $m$ is defined as:
\begin{equation*}
     \cm{X}_{\xpxq}^m\!=\!\{x_i \ | \ s_i \in \cm{R}_{\xpxq} \setminus \cm{O}_{\xpxq}^m, \ v_i \in \cm{V}  \},
\end{equation*}
\noindent where $\cm{O}_{\xpxq}^m\subset \cm{R}_{\xpxq}$ is the set of forbidden positions.
\end{definition} 
\noindent 
In addition, the agents may encounter moving obstacles while executing their tasks, e.g., pedestrians crossing if they represent self-driving cars. Finally, let us introduce the following assumptions: 
\vspace{-4pt}
\begin{assumption}\label{as:mode_m}
The mode $m$ is known \textcolor{black}{at time 0 of each task} by the agents and does not change during its execution, whereas the moving obstacles are discovered while the agents perform the tasks.
\end{assumption}

\begin{assumption}\label{as:initial_solution}
The agents know an initial feasible state trajectory $\bm{\chi}_{\xpxq}^{m}$ and its corresponding input sequence $\bm{\upsilon}_{\xpxq}^{m}$ to complete any task $\xpxq$ in any of its possible modes $m$ \textcolor{black}{without considering moving obstacles}. 
\end{assumption}

\begin{assumption}\label{as:path_mo}\textcolor{black}{
The moving obstacles move along an assigned path, e.g., a crosswalk in the case of pedestrians.}
\end{assumption}

\subsection{Control goal}\label{sec:control_goal}
The control goal is to design the agents'~MPC problems such that they perform optimally any task $p\rightarrow q$ in any of its modes of operation $m$, and can avoid collisions with moving obstacles. In this regard, the stage performance cost for any task with target $\xq$ is assumed to be defined as
\begin{equation} \label{eq:stage_cost}
h(x_i(k), u_i(k),\xq)=   \Vert x_i(k) \!-\! \xq \Vert_{Q}^2\!+\!\Vert u_i(k) \Vert_{R}^2,
\end{equation}
\noindent where $Q$ and $R$ are positive definite matrices. Therefore,~$h(\cdot)$ satisfies $h(\xq, 0,\xq)=0$ and  $h(x_i(k), u_i(k),\xq)>0$ for any $x_i(k)\neq\xq$, $u_i(k) \neq 0$.

\section{Collaborative trajectory optimization based on LMPC}\label{sec:collab_LMPC}

In this paper, the inputs optimization is based on the LMPC presented in~\cite{rosolia2017learning}. 
This controller \textit{learns} its optimal terminal constraints and cost to admissibly perform a task. The formulation is characterized by the following: 
\begin{itemize}
    \item[(i)] The terminal set is a \textit{sample safe set} that is built up from previous successful trajectories. 
    \item[(ii)] The terminal cost weights the cost-to-go from the terminal state to the desired target according to the data collected in previous repetitions of the tasks. 
\end{itemize}

Let us define the sequences of visited states and implemented inputs in the $j$-th task performed by agent~$i$ as
\begin{equation}\label{eq:tray_x_u_ij}
    \mathbf{x}_i^j\!=\!\left[ x_i(t_i^j), \hdots, x_i(T_i^j) \right]\!, \ \  \mathbf{u}_i^j\!=\!\left[ u_i(t_i^j), \hdots, u_i(T_i^j) \right]\!, 
\end{equation}
\noindent where $t_i^j$ and $T_i^j$ represent respectively the instants in which~$i$ started and finished its task number $j$. 
 Likewise, let $m_i^j$ be the mode  in which~$i$ executed its $j$-th task, and consider set
\begin{equation}
   \cm{M}_{i,\xpxq}^m\!=\! \left\lbrace j \ |  \ x_i(t_i^j)\!=\!\xp, \ x_i(T_i^j)\!=\!\xq, \ m_i^j\!=\!m \right\rbrace\!,
\end{equation}
\noindent which contains the indexes of the tasks in which agent~$i$ performed $\xpxq$ in mode $m$.  

\subsection{Cloud storage}

  As shown in Fig.~\ref{fig:architecture}, the proposed approach integrates the set of local agents, a coordinator, and a cloud database.  The cloud stores the data that allows the agents to build up the terminal safe set and terminal cost as described in \cite{rosolia2017learning}.  In this regard, consider the following assumption:
  \vspace{2pt}
   \begin{assumption}
  All agents $i \in \cm{N}$ collect and upload to the cloud their sequences of visited states and incurred costs when performing their tasks. 
  \end{assumption}

 Regarding the terminal set, any agent $i$ can derive a \textit{sample safe set} $\cm{SS}_{i,\xpxq}^m$  for task $\xpxq$ in mode~$m$ by collecting  all its visited states when executing it, i.e.\footnote{For the sake of simplicity, we have omitted the time index in $\cm{SS}_{i,\xpxq}^m$, but note that it represents the safe set according to the trajectories realized up to current instant $k$. 
 },
\begin{equation}\label{eq:SS_xpxq}
\begin{split}
        &\cm{SS}_{i,\xpxq}^m= \left\lbrace \bigcup_{j\in \cm{M}_{i,\xpxq}^m}\bigcup_{t=t_i^j}^{T_i^j} x_i(t)\right\rbrace\!. 
\end{split}
\end{equation}
\noindent  Note that for every $x_i(t) \in \cm{SS}_{i,\xpxq}^m$, with $t\in[t_i^j, T_i^j]$, there exists a sequence of inputs that drives the subsystem state to target $\xq$ satisfying the constraints of mode~$m$, i.e.,~$[u_i(t),  \hdots,  u_i(T_i^j)]$. Accordingly, all states in sequences~$\mathbf{x}_i^j$, for all $j \in \cm{M}_{i,\xpxq}^m$, belong to the maximal control invariant set associated with task $\xpxq$ and mode~$m$.

Also, to define the terminal cost function, we store for all~$x_i(t) \in \cm{SS}_{i,\xpxq}^m$ the corresponding cost-to-go to complete the task, i.e., $F_i(t)=\sum_{k=t}^{T_i^j} h(x_i(k), u_i(k),\xq)$, 
\noindent where $t \in [t_i^j, T_i^j]$ and $j \in \cm{M}_{i,\xpxq}^m$.  

\begin{remark}\label{rem:union_SS}
Since  all the agents are dynamically identical, set~$\cm{SS}_{i,\xpxq}^m$ also contains \textit{safe} states for any other agent~$j \in \cm{N}\setminus\{i\}$. 
Therefore, one can define a \textit{common sample safe set} for all agents as: 
\begin{equation}\label{eq:union_SS}
\cm{SS}_{\xpxq}^m= \left\lbrace \bigcup_{i \in \cm{N}} \cm{SS}_{i,\xpxq}^m\right\rbrace\!.
\end{equation}
For the same reasons, the costs-to-go $F_i(\cdot)$ also provide useful information for all $j \in \cm{N}\setminus\{i\}$. 
\end{remark}

\begin{remark}\label{rem:similarity}
By exploiting the similarity between tasks, it is possible to use data collected in a certain task to learn a different one. For example, consider the system in Fig.~\ref{fig:architecture}
and assume that the $j$-th task performed by agent $i$ was $\xpxq$. 
Additionally, consider a different task $r\rightarrow n$, and assume that roads $\cm{R}_{\xpxq}$ and~$\cm{R}_{r\rightarrow n}$ are parallel and satisfy $\cm{R}_{r\rightarrow n}= \{(s_\nc{x}, s_\nc{y})+\lambda \ : \  (s_\nc{x}, s_\nc{y})  \in  \cm{R}_{\xpxq} \}$, where $\lambda$ is the shift vector. Then, 
trajectories $\n{u}_i^j$ and
\begin{equation*}
    \n{x}_i^j+\lambda = [x_i(t_i^j)+\lambda, \ x_i(t_i^j+1)+\lambda, \hdots, \ x_i(T_i^j)+\lambda]
\end{equation*} 
provide a new candidate solution for task $r\rightarrow n$. One should simply check if it satisfies the constraints of some of the modes of $r\rightarrow n$, and add the new solution to the corresponding safe set if admissible. Note also that if the problems associated with tasks $\xpxq$ and $r\rightarrow n$ are convex, then a convex combination of the trajectories obtained for $\xpxq$ shifted by~$\lambda$ also provides safe points for task $r\rightarrow n$. 
\end{remark}




\subsection{Local controllers}

Consider any agent $i\in \cm{N}$, let $\xpxq$ be its assigned task at time instant $k$, and $m$ be the currently active mode for this task. Then, the input implemented by agent $i$ is computed by solving the following problem:  
\begin{subequations}\label{eq:LMPC_agents}
\begin{align}
      \min_{\vu_i} &\sum_{t=k}^{k+N-1} h(x_i(t|k), u_i(t|k), \xq) \label{eq:obj_func_LMPC_agents} +  P_{\xpxq}^m(x_i(k+N|k)) \nonumber \\[2pt]
     \text{s.t.} \ \ 
    & x_i(k|k) = x_i(k),\\
    & x_i(t+1|k) = A x(t|k) + B u_i(t|k),  \\
    &  x_i(t+1|k) \in  \cm{X}_{\xpxq}^m, \ u_i(t|k) \in \cm{U}, \label{eq:state_constraints_LMPC_agents} \\
    & \forall t \in [k,..., k+N-1], \\
    & x_i(k+N|k) \in \cm{SS}_{\xpxq}^m, \label{eq:term_const_LMPC}
    \end{align}
    \end{subequations}
\noindent where $N$ is the prediction horizon, $\vu_i$ is the sequence of inputs from instant $k$ to $k +N -\!1$, and~$(t|k)$ denotes a prediction for time instant $t$ made at $k$.  Additionally, function $P_{\xpxq}^m(\cdot)$ assigns to every point in $\cm{SS}_{\xpxq}^m$ the minimum cost to complete the task according to the data collected up to instant $k$. 
Note that this value can be computed from the costs realized in previous executions of the tasks.



\subsubsection{\textcolor{black}{Moving obstacles}}

As mentioned in Section~\ref{sec:agents_tasks}, in addition to the modes, the agents may encounter moving obstacles while executing the tasks. In this regard, consider that agent~$i$ is performing task $\xpxq$ in mode~$m$ by solving~\eqref{eq:LMPC_agents}, and that it detects a moving obstacle at instant~$k$. Then, it switches to the following modified problem: 
 \\
 \vspace{-0.7cm}
 \\
\begin{subequations}\label{eq:LMPC_agents_transition}
\begin{align}
 \min_{\vu_i}& \sum_{t=k}^{k+N-1} h(x_i(t|k), u_i(t|k), \xq)  \nonumber\\
     \text{s.t.} \ \ \ 
    & x_i(k|k) = x_i(k),\\
    & x_i(t+1|k) = A x_i(t|k) + B u_i(t|k),  \\
    & x_i(t+1|k) \in \cm{X}_{\xpxq}^{\nc{bo},m}, \ u_i(t|k) \in \cm{U}, \label{eq:state_const_pedestrian}\\
    & s_i(k+N|k) \in \text{Proy}_{s}(\cm{SS}_{\xpxq}^m),  \label{eq:term_const_LMPC_pedestrian} \\
   & \forall t \in [k,..., k+N-1], 
    \end{align}
    \end{subequations}
    \noindent where $\text{Proy}_{s}(\cdot)$ denotes the projection onto the components associated with the position. 
    Also,~$\cm{X}_{\xpxq}^{\nc{bo},m}$ represents the subset of~$\cm{X}_{\xpxq}^{m}$ such that~$\text{Proy}_{s}(\cm{X}_{\xpxq}^{\nc{bo},m})$ is the area of the road before the \textcolor{black}{path along which the obstacle moves (see Assumption~\ref{as:path_mo})}. Note that the speed is not constrained in~\eqref{eq:term_const_LMPC_pedestrian} since agent $i$ may need to slow down to remain in this area. Once the \textcolor{black}{moving obstacle is no longer detected}, it starts solving again \eqref{eq:LMPC_agents} to continue traveling to target~$\xq$.  




\subsection{Algorithm}

 Algorithm~\ref{alg:local_agent} provides the pseudo-code of the procedure followed by each agent $i\in\cm{N}$ in the case without \textcolor{black}{moving obstacles}. This procedure can be run in parallel by all agents in the system. Note that the sharing of information is involved in the download/upload of data from/to the cloud, which only occurs at the beginning and end of the tasks. That is, there is no agent-to-agent communication, and they may also access the cloud at different time instants. In addition, note that the goal of the coordinator is only to assign the tasks to the agents and inform them of the mode of operation.\footnote{The analysis and optimization of the tasks' assignments have been left out of the scope of this paper.}
 
%


\begin{algorithm}[h]
\caption{}
Initialize $\cm{SS}_{\xpxq}^{m}$ for all tasks and modes using initial solution $\bm{\chi}_{\xpxq}^{m}$, and define~$\cm{N}_\nc{free}$  as the set of agents that are ready to be assigned a task. Also,  consider some agent~$i \in \cm{N}_\nc{free}$, let its  state be $\xp \in \cm{X}$, and $j$ its tasks' counter. Then, at each step~$k$, consider the following:
\begin{algorithmic}[1]
\IF{$i\in \cm{N}_\nc{free}$}
\STATE The coordinator assigns a new target state $\xq \in \cm{\hat{X}}$ to~$i$, with $q\neq p$, and informs it about the mode $m$.
\STATE Agent $i$ accesses the cloud to get safe set $\cm{SS}_{\xpxq}^{m}$, and saves the starting instant of the task, i.e.,~$t_i^j\!=\!k$. 
\STATE The coordinator updates $\cm{N}_\nc{free} = \cm{N}_\nc{free} \setminus \{ i \}$.
\ENDIF
\STATE Agent $i$ finds its input by solving \eqref{eq:LMPC_agents} for its assigned task,  and  updates its state.
\IF{ the task is completed}
\STATE Agent $i$ uploads to the cloud its realized states $x_i(t)$, for $t\in[t_i^j,k]$, and they are added to $\cm{SS}_{\xpxq}^{m}$.
\STATE  The costs-to-go $F_i(t)$, for $t\in[t_i^j,k]$, are also computed and uploaded to the cloud.
\STATE It is checked whether the new data can be exploited for other tasks. 
\STATE Set $j= j+1$, $\cm{N}_\nc{free}= \cm{N}_\nc{free} \cup \{i\}$, and $p=q$.
\ENDIF
\end{algorithmic}
\label{alg:local_agent}
\end{algorithm}

 \begin{remark}
 Detecting a moving obstacle only implies a change in Steps~6,~8 and~9 of Algorithm~\ref{alg:local_agent}. In particular, instead of solving problem~\eqref{eq:LMPC_agents}, the agents would use~\eqref{eq:LMPC_agents_transition} until it stops being detected.  Likewise, the data altered by the presence of moving obstacles should not be uploaded to the cloud for its use to define the terminal cost and set of \eqref{eq:LMPC_agents}.
Finally, note that there is no change of constraints during the prediction horizon, that is, the agents either solve~\eqref{eq:LMPC_agents} or~\eqref{eq:LMPC_agents_transition}. 
 \end{remark} 
 



\section{Theoretical properties}

This section describes the theoretical properties of the proposed controller. Let $[x_i^*(k|k), x_i^*(k+1|k), ..., x_i^*(k+N|k)]$ be the optimal state sequence computed by agent $i$ at time instant~$k$, and $[s_i^*(k|k), s_i^*(k+1|k), ..., s_i^*(k+N|k)]$ the associated sequence of positions. Similarly, for state sequence $[x_i(k), x_i(k+1), ..., x_i(k+t)]$ with $t>1$, vector $[s_i(k), s_i(k+1), ..., s_i(k+t)]$ is the corresponding position sequence.  Finally, let us introduce the following assumptions: 





\begin{assumption}\label{ass:Delta_pedestrians}
Any agent $i\in \cm{N}$ is able to detect moving obstacles before the Euclidean distance between $i$ and the obstacle is lower than or equal to $\Delta$, with $\Delta$ being the maximum distance the agents can travel in $N$ steps. 
\end{assumption}

\begin{assumption}\label{ass:more_steps}
Consider a sequence $[x_i(k), x_i(k+1), \hdots, x_i(k+t)]$ such that $x_i(k+n) \in \mathcal{X}_{\xpxq}^{\nc{bo},m}\subset \mathcal{X}_{\xpxq}^m$ for all $n=0, ...,t$. Then,  it is possible to travel from position~$s_i(k)$ to $s_i(k+t)$ in $T\geq t$ time steps following a state sequence that also remains in $\mathcal{X}_{\xpxq}^{\nc{bo},m}$.
\end{assumption}

\begin{assumption}\label{ass:position} 
Let $x_i(k)$ be such that $s_i(k) \in \text{Proy}_{s}(\cm{SS}_{\xpxq}^m)$. Then, there exists an admissible sequence of inputs such that $x_i(k+t) \in \cm{SS}_{\xpxq}^m$ for some $t\geq 0$. 
\end{assumption}

\noindent Then, the following theorems hold.

\begin{thm}[Recursive feasibility]
  All agents $i \in \cm{N}$ can find a feasible solution of their optimization problems at all instants $k \geq 0$. 
\end{thm}

\textit{Proof.} This proof is divided in two parts. First, we focus on the case in which  there are no moving obstacles.

Consider that at time instant $k$ agent $i\in \cm{N}$ is assigned task~$\xpxq$ in mode $m$. Also, assume that the latter was also the $z-$th task previously completed by agent~$l$.  Then, the first~$N$ states of trajectory~$\n{x}_l^z$  and its associated input sequence (see \eqref{eq:tray_x_u_ij}) provide a feasible solution of problem \eqref{eq:LMPC_agents} for agent~$i$ at~$k$. 
Also, if no agent has performed task $\xpxq$ in mode $m$ before, the similar result can be derived by using the initial solution~$(\bm{\chi}_{\xpxq}^m, \bm{\upsilon}_{\xpxq}^m)$ (see Assumption~\ref{as:initial_solution}).

 Now let us move to instant $k+1$. Given~\eqref{eq:term_const_LMPC}, $x_i^*(k+N|k)$ belongs in an admissible state trajectory to complete the task.  Let $x_i^*(k+N|k)$ correspond to the state at instant $t$ of some agent $l$, and take the following candidate solution:
\begin{equation}\label{eq:opt_sol_ext}\resizebox{.89\hsize}{!}{$
\begin{split}
     &\begin{bmatrix} x_i^*(k+1|k) & x_i^*(k+2|k) & \hdots & x_i^*(k+N|k) & x_l(t+1) \end{bmatrix}\!, \\
     &\begin{bmatrix} u_i^*(k+1|k) & u_i^*(k+2|k) & \hdots & u^*_i(k+N-1|k) & u_l(t) \end{bmatrix}\!.
    \end{split}$}
\end{equation}
\noindent Given that $x_i(k+1)=x_i^*(k+1|k)$, that sequences computed at $k$ are admissible,  and that $x_l(t+1) \in \cm{SS}_{\xpxq}^{m}$, we have that~\eqref{eq:opt_sol_ext} provides a feasible solution of problem~\eqref{eq:LMPC_agents} at time instant~$k+1$. By induction, recursive feasibility is guaranteed. 

The recursive feasibility proof when moving obstacles come into play is based on Assumptions \ref{ass:Delta_pedestrians}, \ref{ass:more_steps} and \ref{ass:position}. Consider that agent $i$ solves \eqref{eq:LMPC_agents} at instant $k+\tau$, with $\tau> 0$. If at~$k+\tau+1$ it detects a moving obstacle, it switches to problem~\eqref{eq:LMPC_agents_transition}. From the solution of \eqref{eq:LMPC_agents} at~$k+\tau$, and considering Assumption~\ref{ass:Delta_pedestrians}, we have
\begin{subequations}\label{eq:kplustau}
\begin{align}
        &x_i(k+\tau+1)=x_i^*(k+\tau+1|k+\tau), \label{eq:x_ini}
        \\ 
        &x_i^*(k+\tau+t|k+\tau) \in \cm{X}_{\xpxq}^{\nc{bo},m}, \ \  \forall t=1, ..., N,  \label{eq:x_bo}
        \\ &x_i^*(k+\tau+N|k+\tau) \in \cm{SS}_{\xpxq}^{m}.\label{eq:terminal_state_tau}
\end{align}
\end{subequations}
Given \eqref{eq:x_ini}, \eqref{eq:x_bo}, and Assumption~\ref{ass:more_steps}, it is possible to find a sequence of inputs to go from $s_i(k+\tau+1)$ to $s_i^*(k+\tau+N|k+\tau)$ in~$N$ steps while remaining in the area of the road before the path along which the obstacle moves. Also, from \eqref{eq:terminal_state_tau}, we have that $s_i^*(k\!+\!\tau\!+\!N|k+\tau) \in \text{Proy}_{s}(\cm{SS}_{\xpxq}^m)$, hence the latter would provide a feasible solution of problem~\eqref{eq:LMPC_agents_transition}. Finally, assume that the \textcolor{black}{moving obstacle stops being detected} at~$k+\Gamma+1$, with $\Gamma>\tau$, and, therefore,  agent~$i$ switches back to problem~\eqref{eq:LMPC_agents}. The only constraint that may compromise feasibility is~\eqref{eq:term_const_LMPC}. Under Assumption~\ref{ass:position}, agent~$i$ can find an admissible trajectory to go from  terminal state  $x_i^*(k+\Gamma+N|k+\Gamma)$ to the safe set, e.g.,  by traveling to positions in $\text{Proy}_{s}(\cm{SS}_{\xpxq}^m)$ while progressively adjusting its speed. Likewise,  the solution at $k+\Gamma$ provides an admissible input sequence to get from $x_i(k+\Gamma+1)$ to~$x_i^*(k+\Gamma+N|k+\Gamma)$. By aggregating both input sequences, a feasible solution  of problem \eqref{eq:LMPC_agents} at $k+\Gamma+1$ is defined for a certain horizon. Note that one could recalculate \eqref{eq:LMPC_agents}  using a larger $N$ if it is necessary.  $\qed$

\begin{thm}[Stability for each task]
For any task $\xpxq$ and mode $m$, the equilibrium point $\xq$ is asymptotically stable.
\end{thm}
\noindent In the no \textcolor{black}{moving obstacles} case,  asymptotic convergence to  the target  is proven by showing that the optimal cost is a Lyapunov function for equilibrium point~$\xq$, with the cost function being the objective in \eqref{eq:LMPC_agents}. 
The latter is proven by following the same steps as in \cite[Thm.~1]{rosolia2017learning}. 
On the other hand, if agent $i$ detects a \textcolor{black}{moving obstacle}, the value of the cost function in \eqref{eq:LMPC_agents} may increase while its inputs are computed by solving \eqref{eq:LMPC_agents_transition}. Nonetheless, assuming that the detection of the latter is sufficiently spaced out in time, asymptotic stability of the target points can also be proven. In particular, consider that the last \textcolor{black}{moving obstacle stops being detected}  at instant~$k+\Gamma+1$, then, from that instant until the end of the task, agent $i$ will always use \eqref{eq:LMPC_agents}. During this period, the decreasing evolution of the cost function will be guaranteed, and we can use the same reasoning as in \cite[Thm.~1]{rosolia2017learning} to conclude that $i$ will converge to target $\xq$.

\begin{thm}[Non-increasing costs]\label{th:non_increasing}
The costs of performing any task $\xpxq$ in mode $m$ without moving obstacles decrease with the number of repetitions, regardless of the agents that perform it. 
\end{thm}

\noindent The proof of Theorem~\ref{th:non_increasing} can be easily derived from \cite[Thm.~2]{rosolia2017learning}. In particular, using \cite[Thm.~2]{rosolia2017learning}, it is straightforward to prove that the cost of executing a given task decreases as some agent repeats it. Since in the proposed scheme all agents are identical and share their collected data, this non-increasing property holds equally. 

\vspace{6pt}

Considering the theorems above and assuming that the agents converge to a steady-state trajectory for each $\xpxq$ and mode $m$, it is possible to find a link with the associated infinite-horizon problem~\cite[Thm.~3]{rosolia2017learning}. In particular,  the  steady-state  trajectory is proven to match the optimal solution of a finite-time approximation of~the latter if it is convex. 

\section{Simulation results}

\begin{figure}[t]
    \centering
    \includegraphics[scale=0.5,  trim={1.5cm 13.5cm 9cm 9.1cm},clip]{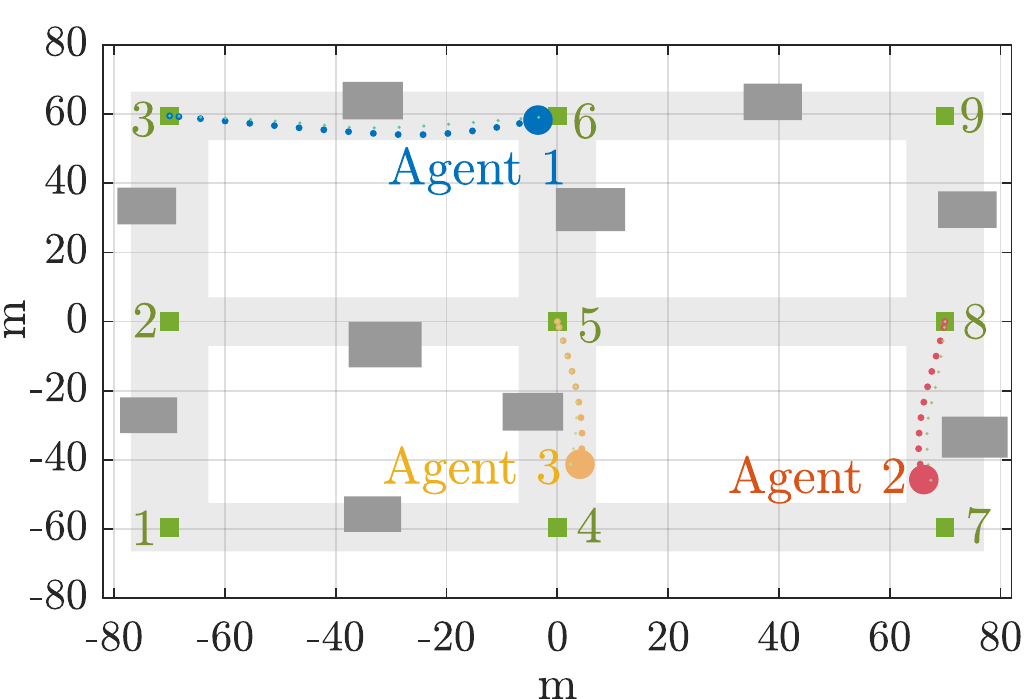}
    \vspace{-0.2cm}
    \caption{Layout of the plant. The dark grey areas indicate the forbidden zones, which vary in time and determine the modes of operation.} 
    \label{fig:mapa}
\end{figure}

Consider a system where 3 agents travel between a set of predefined locations by using the cloud-based LMPC described above. Also, assume that the plant layout is as shown in Fig.~\ref{fig:mapa}, where the intersections correspond to the target locations. 
The conditions for performing the tasks vary between two modes of operation, i.e., $m=\{0, 1\}$. 
 These modes change independently for each of the tasks since the conditions in each road can change at different time steps.
In addition, 
we consider that at the middle of each road there is a crosswalk where the agents find pedestrians, which are handled as moving obstacles, with a probability of 10\%.


The parameters of the simulation are: $v_\nc{max}=3$m/s, $a_\nc{max}=1.5$m/s$^2$, $N=4$, $Q=0.01 I_4$, and $R=0.5 I_2$,
where $I_4$ and $I_2$ represent respectively the identity matrices of dimensions $4 \times 4$ and $2 \times 2$. The discrete model of the agents was obtained by discretizing \eqref{eq:robots_model} with a sample time of 1.5s.    Also, the condition to terminate any task with target~$\xq$ is defined as  $\Vert s_i-\hat{s}^q \Vert_{\infty}\leq 0.01$ and $\Vert v_i-v^q \Vert_{\infty} \leq 0.001$ for all $i$. Finally, the safe sets were computed as in~\eqref{eq:union_SS}, and we exploited the parallelism between tasks (see Remark~\ref{rem:similarity}). The results obtained are summarized below\footnote{To reduce the computational complexity, we have used the convex approximation of the safe sets when solving \eqref{eq:LMPC_agents} (see 
 \cite{rosolia2017efficient}).}\footnote{To implement the state constraints in \eqref{eq:state_constraints_LMPC_agents} we have imposed $\Vert x_i(t|k) - c_{\xpxq}^m \Vert\geq D_{\xpxq}^m$, for all $t\!=\!k+1,...,k\!+\!N$, where $c_{\xpxq}^m$ is the Chebyshev center of the forbidden area and $D_{\xpxq}^m$ a properly defined distance.}.

\begin{figure}[t!]
    \centering \hspace{-1cm}
    \includegraphics[scale=0.55,  trim={2.6cm 10.1cm 0 10cm},clip]{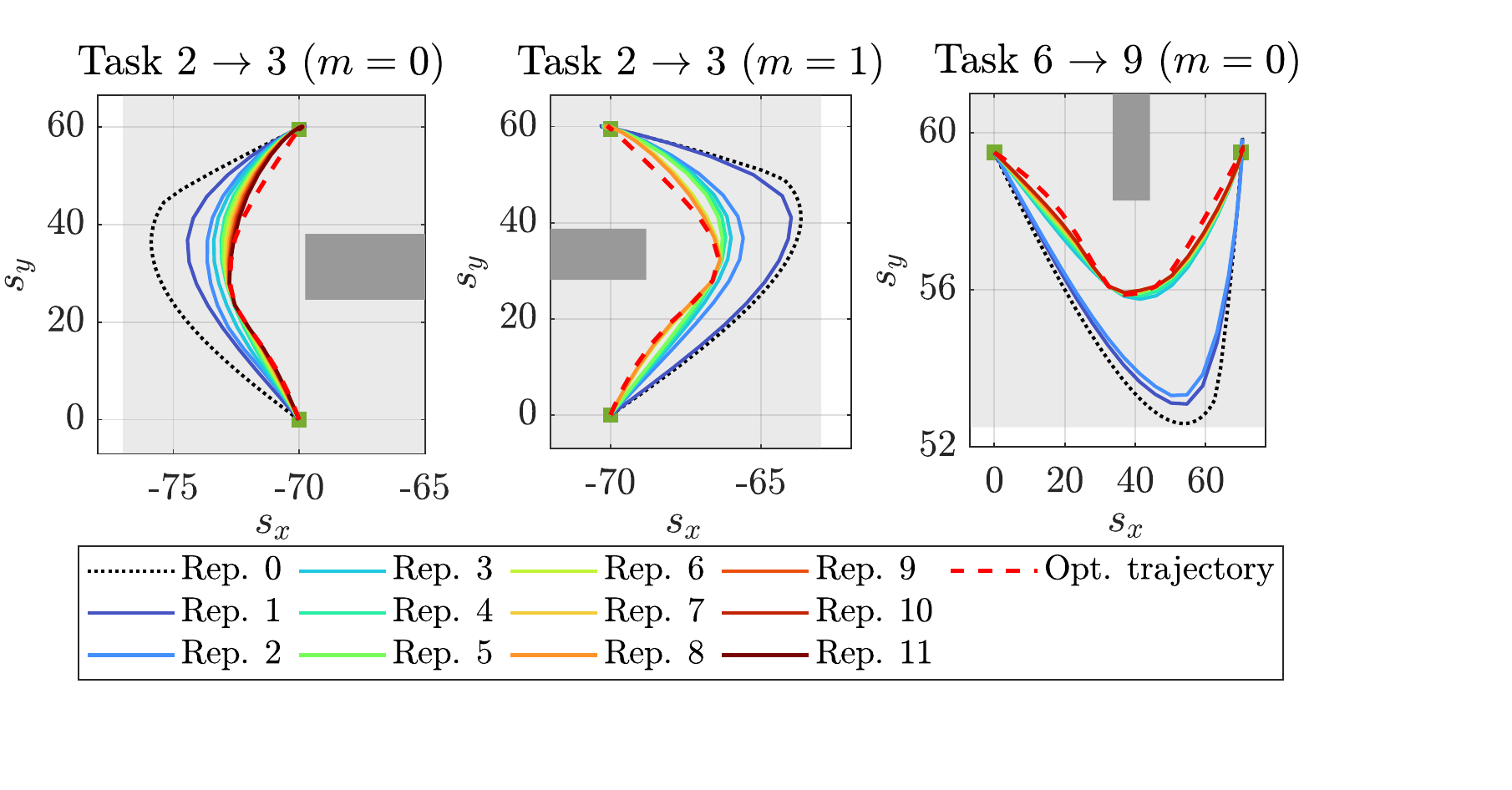}
    \vspace{-1.2cm}
    \caption{Agents' position trajectories in each repetition of different tasks. As the agents get and share more data, their trajectories are progressively shifted from the initial feasible solution towards the optimal one.}
    \label{fig:trajectories}
\end{figure}

Fig.~\ref{fig:trajectories} illustrates the evolution of the agents' position for different tasks. 
The significant difference between consecutive repetitions that can particularly be seen in the case of $2\rightarrow3$ and $6\rightarrow9$ reflects the effect of using data from parallel tasks to accelerate the learning process. That is,  between two executions of the same task, the agents may perform others providing useful data for the one at issue. Notice that given footnote 5, the constraints consider a circular area surrounding the dark grey zones as forbidden. Also, note that for each $\xpxq$ and mode $m$, the \textit{optimal} trajectory is considered to be the one obtained by optimizing the sum of stage cost \eqref{eq:stage_cost} for a very long prediction horizon subject to the corresponding constraints.

\begin{figure}[t!]
    \centering
    \includegraphics[scale=0.5,  trim={2.3cm 12.4cm 0 9.1cm},clip]{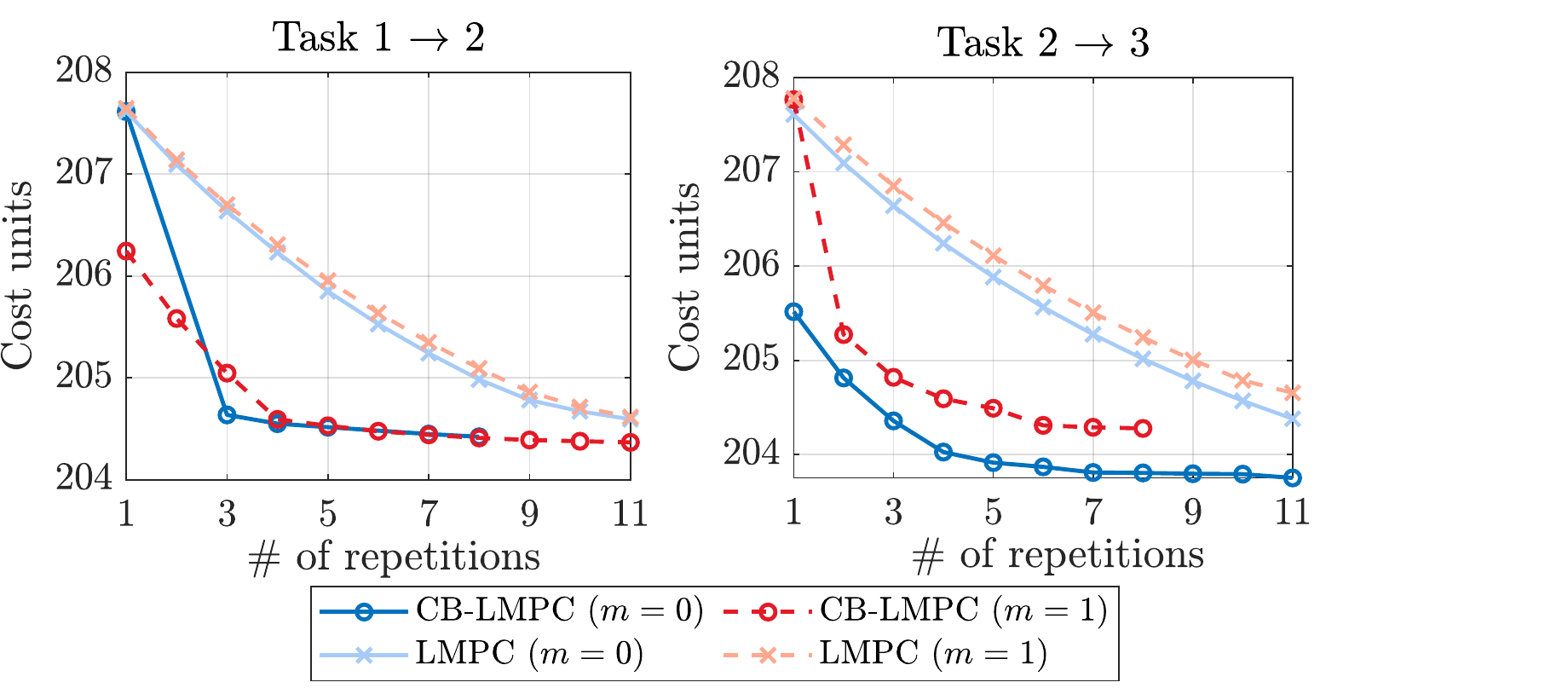}
    \vspace{-0.6cm}
    \caption{Evolution of the costs of performing different tasks  with the proposed cloud-based approach and when using the original LMPC formulation~\cite{rosolia2017learning}. The number of repetitions of each task varies because the coordinator assigned them randomly.  }
    \label{fig:cost_tasks}
\end{figure}

\begin{figure}[t]
    \centering
    \includegraphics[scale=0.57,  trim={3.0cm 11.3cm 0 12.2cm},clip]{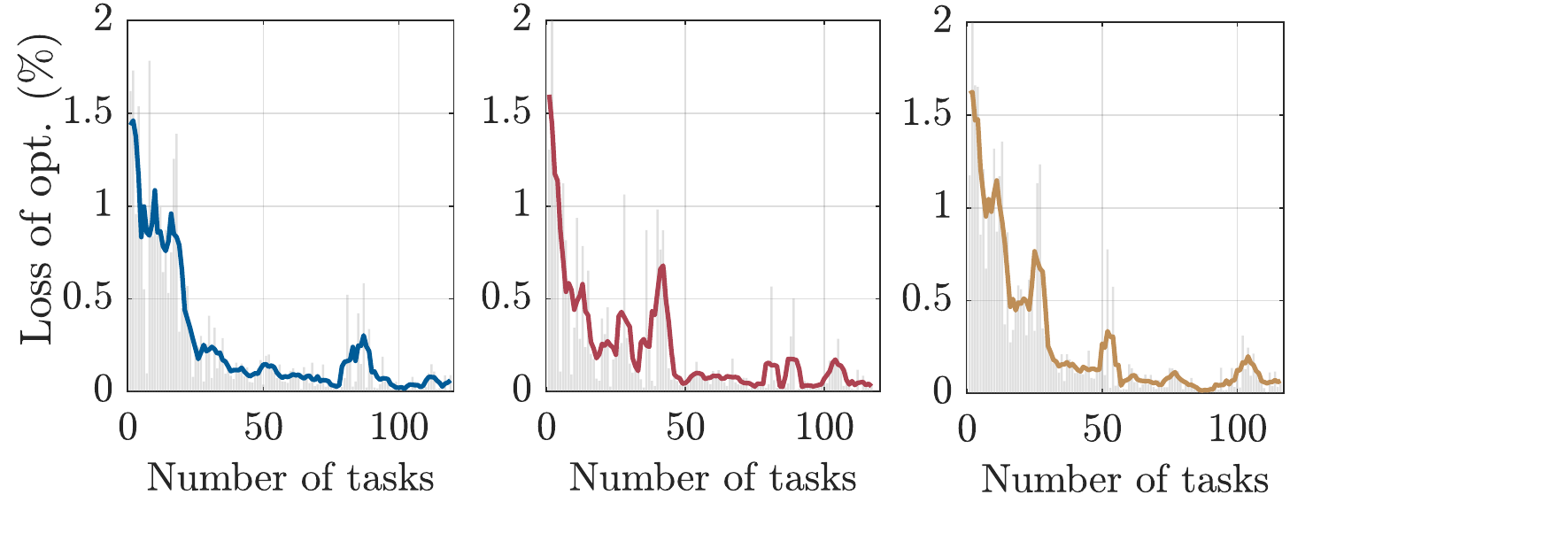}
    \vspace{-0.8cm}
    \caption{Losses of optimality obtained by agents 1 (left), 2 (middle), and~3 (right) in their tasks.  The lines show the corresponding
moving averages considering a sliding window of 5 tasks. }
    \label{fig:cost_agents_and_numtasks}
\end{figure}

 Fig.~\ref{fig:cost_tasks} compares the cost of executing two different tasks with the results obtained using the original~LMPC in~\cite{rosolia2017learning}, i.e., the same formulation but without considering the storage cloud and the exploitation of the similarity between tasks. For the sake of convenience, we use CB-LMPC to refer to the proposed cloud-based approach, and simply LMPC to refer to~\cite{rosolia2017learning}. 
   As can be seen,  the costs follow a decreasing trend and tend to stabilize at the same value for a given~$p$,~$q$, and~$m$. 
Nevertheless, the CB-LMPC allows us to obtain lower costs during the learning process, and to converge to a solution in a reduced number of repetitions. In addition, Fig.~\ref{fig:cost_agents_and_numtasks} illustrates the loss of optimality incurred by the agents in their tasks with the CB-LMPC.   Note that there are 48 possible tasks (12 roads, two ways, and two modes), and that each agent performed about~130. Therefore, on average, they repeated less than~3 times each possible task. 
It should also be remarked that Fig.~\ref{fig:cost_tasks} and~\ref{fig:cost_agents_and_numtasks} focus on tasks not affected by pedestrians, since they are the ones that should lead to non-increasing costs. 

\begin{figure}[t]
    \centering
    \includegraphics[scale=0.5,  trim={2.2cm 11.5cm 0 11.2cm},clip]{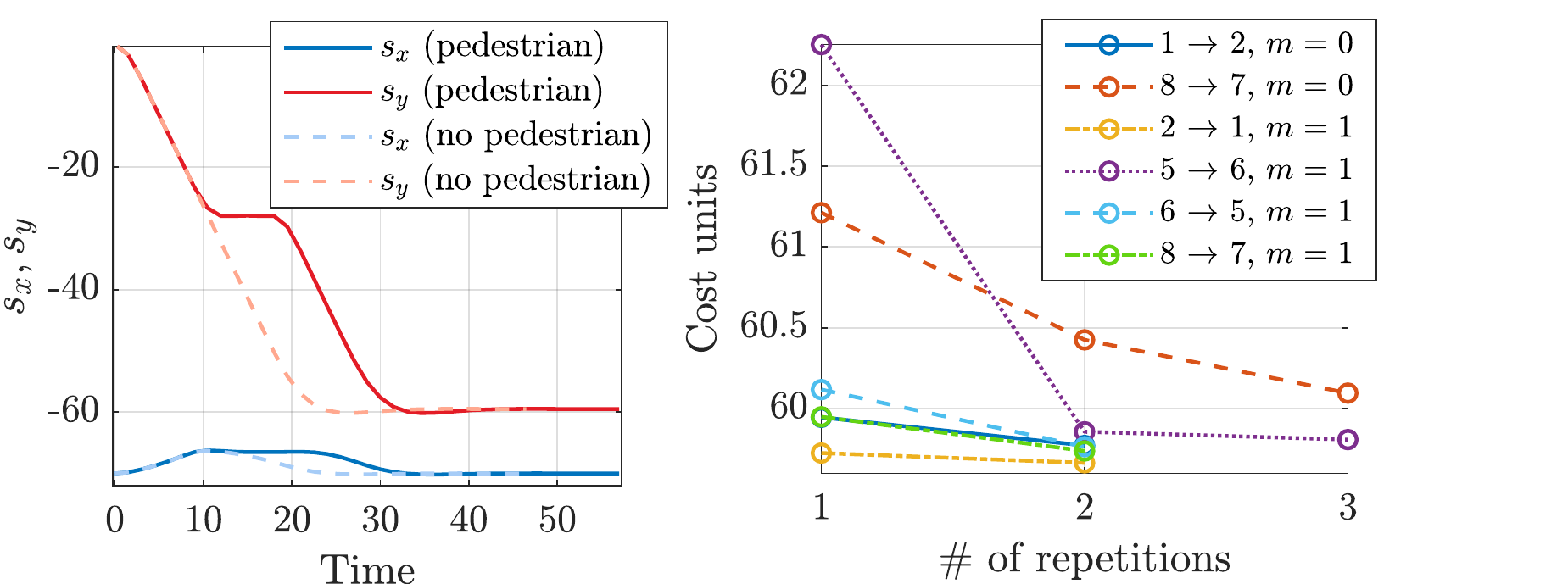}
    \vspace{-0.6cm}
    \caption{Deviation of the position evolution (left) and extra costs (right) caused by the presence of pedestrians. The left plot shows the position of agent 2 when it encounters a pedestrian while performing task $2 \rightarrow 1$, and the result at convergence for the case without pedestrians. }
    \label{fig:pedestrian}
\end{figure}

Finally, Fig.~\ref{fig:pedestrian} (left) illustrates the effect of detecting pedestrians on the agents' position. As can be seen, it remains approximately constant between instants~10s and~18s, which corresponds to the  waiting period while the pedestrian crosses. 
Also, Fig.~\ref{fig:pedestrian}~(right) illustrates the excess of costs over the optimal one in tasks affected by pedestrians more than once. 
It can be seen how the data collected to optimize performance without pedestrians helps indirectly to reduce the costs in this~case.

\section{Conclusions}

In this paper, a LMPC for multi-agent systems that perform repetitive tasks is presented.  The agents collect their state trajectories and incurred costs, and upload them to a common cloud to help their own and their peers' performance. 
Moreover, we have distinguished different modes of execution of these tasks, which modify the state constraints to deal with static obstacles. Also, we have included a strategy for reacting to \textcolor{black}{moving obstacles},  which is guaranteed not to compromise recursive feasibility under the introduced assumptions. The data in the cloud are in turn analyzed to check if some trajectories can be concurrently exploited for more than one task. 
Our results show that the presented LMPC can notably improve the learning rate in a context where a significant number of tasks~are~alike.

Future work will extend the proposed architecture to systems where the agents' dynamics and constraints are not identical. Additionally, we will integrate the possible presence of moving obstacles directly in the LMPC formulation to guarantee non-increasing costs also in this case.

\addtolength{\textheight}{-12cm}   





\bibliographystyle{IEEEtran}
\bibliography{biblio.bib}

\end{document}